# Hyperspectral optical diffraction tomography

**JaeHwang Jung, Kyoohyun Kim, Jonghee Yoon, and YongKeun Park***

*Department of Physics, Korea Advanced Institutes of Science and Technology, Daejeon 305-701, Republic of Korea*
*\*yk.park@kaist.ac.kr*

Here, we present a novel microscopic technique for measuring wavelength-dependent three-dimensional (3-D) distributions of the refractive indices (RIs) of microscopic samples in the visible wavelengths. Employing 3-D quantitative phase microscopy techniques with a wavelength-swept source, 3-D RI tomograms were obtained in the range of 450 – 700 nm with a spectral resolution of a few nanometers. The capability of the technique was demonstrated by measuring the hyperspectral 3-D RI tomograms of polystyrene beads, human red blood cells, and hepatocytes. The results demonstrate the potential for label-free molecular specific 3-D tomography of biological samples.

## 1. Introduction

Three-dimensional (3-D) molecular imaging has a crucial role in modern biology and medicine. As the complexity of cellular biology exponentially increases, multiplexing becomes necessary with multi- or hyperspectral imaging and 3-D tomographic imaging capabilities [1]. Conventional hyperspectral 3-D microscopic techniques based on confocal scanning fluorescent microscopy and multiphoton microscopy with the use of multiplexed fluorescent proteins or quantum dots provide high molecular specificity [2]. However, invasive procedures such as chemical stanings or genetic modifications are required for the use exogenous labeling agents which may interfere with intrinsic biological activities [3], and limited numbers of fluorescent agents can be used simultaneously due to the broad excitation and emission spectra of fluorescence agents.

Recently, several 3-D molecular imaging techniques exploiting intrinsic contrasts have been introduced, including molecular imaging spectroscopic optical coherence tomography [4] and 3-D coherent anti-stokes Raman scattering microscopy [5]. However, these techniques are limited to specific molecules exhibiting distinct absorption or vibration spectra. Alternatively, optical dispersion, otherwise known as wavelength-dependent refractive index (RI), has been exploited using quantitative phase imaging techniques [6, 7] for this purpose. Several approaches have been reported for measuring the RI dispersion of biological cells and tissues [8-16].

However, no existing techniques can measure RI dispersion tomography; the established spectroscopic phase imaging techniques measure 2-D images, in which the RI values are coupled with the thickness of a sample as an optical phase delay. So far, although various 3-D quantitative phase imaging techniques have been developed for visualizing 3-D RI tomograms of cells [7, 17-21] and utilized for the study of cell pathophysiology [16, 22-28], wavelength-dependent 3-D RI tomography of cells has not been demonstrated before. Measurements of 3-D RI tomograms of cells at various wavelengths would have much to offer the fields of cell biology and medicine with its non-invasiveness, quantitative imaging capability, as well as molecular information via optical dispersion [29].

Here, we present a novel approach, referred to as hyperspectral optical diffraction tomography (HS-ODT), which measures the 3-D RI tomograms of microscopic samples at multiple wavelengths ranging from 450 to 700 nm with a diffraction-limited spatial resolution. We integrate a custom-made wavelength-swept source for hyperspectral illumination [10] and common-path diffraction optical tomography for acquiring optical field information with multiple illumination angles [30]. The capability of the HS-ODT is demonstrated by measuring the 3-D RI tomograms of microspheres, human RBCs, and hepatocytes at multiple wavelengths in the visible range.

## 2. Method and results

The HS-ODT setup consists of two parts: a wavelength-swept source and common-path interferometric microscopy shown in the Fig. 1(a). The beam from a super-continuum laser (SuperK COMPACT, NKT Photonics A/S, Denmark) is dispersed using a prism (N-SF11 prism, PS853, Thorlabs Inc. USA), and then, a specific wavelength is filtered by using a pinhole and rotating a single-axis galvano mirror (GM1, GVS011/M, Thorlabs Inc.). The wavelength $\lambda$ of the illumination beam is scanned in the range of 450 – 700 nm with a spectral bandwidth (standard deviation of a Gaussian distribution) of a few nanometers: 1.2 nm at $\lambda$ = 465 nm to 5.9 nm at $\lambda$ =660 nm [Fig. 1(b)].

To measure the hyperspectral tomograms of a sample at a specific wavelength, multiple 2-D optical fields of a sample were obtained for various angles of illumination for each wavelength of the illumination. The incident angle of the illumination beam is precisely controlled by the dual-axis rotating mirror (GM2, GVS012/M, Thorlabs Inc.) located at the plane conjugated to the sample and projected onto a sample by a condenser lens [PLAPON 60×, air immersion, numerical aperture (NA) = 0.9]. The diffracted beam from the sample is collected by a high-NA objective lens (PLAPON 60×, oil immersion, NA = 1.42) and tube lens ($f$ = 180 mm). The tilted angle made by GM2 is compensated by synchronizing with a dual-axis rotating mirror (GM3, GVS012/M, Thorlabs Inc.) so that the optical axis does not change after GM3, regardless of the incident angle. Then, the optical fields of the sample are measured at sCMOS plane (Neo 5.5, 6.5 μm, 100 fps at 5.5 Megapixels, Andor Technology) by interfering the sample beam and the spatially filtered reference beam, using the common-path geometric interferometer consisting of the grating (#46-072, 92 grooves/mm, Edmund Optics Inc.) and a 4-$f$ telescopic imaging system with a spatial filter (20 μm diameter, customized). Groove density of the grating and the size of the pinhole are determined to maximize energy efficiency of the beam and to utilize the full resolution given by the objective lens. Detailed information on the wavelength sweeping source, common-path interferometric microscopy, and field measurement techniques can be found in the Refss [10, 30-32].

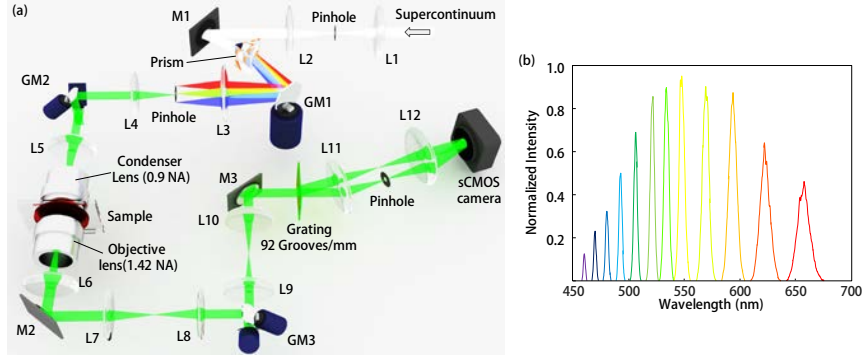

Fig. 1. Experimental setup and spectra of the illumination of HS-ODT. (a) The wavelength of the illumination beam is controlled using a galvamomirror, GM1. The illumination angle impinging onto a sample is scanned by GM2, which is then descanned by GM3. L1-12, lenses; M1-3, mirrors; GM1, single-axis rotating mirror; GM2-3, dual-axis rotating mirrors. (b) Spectral density of the illumination which covers the wavelength range from 450 to 700 nm. Because a prism is used as a dispersion material, bandwidths of individual wavelengths gradually vary from 1.2 nm at $\lambda$ = 465 nm to 5.9 nm at $\lambda$ = 660 nm.

To reconstruct a 3-D RI tomogram of a sample, 300 optical fields from various incident angles were measured. The illumination beam was spirally scanned which sparsely spans the NA of the condenser lens. The time interval between each illumination angle was synchronized to the camera. Once the recording of a series of holograms at a single wavelength is complete, the illumination wavelength is changed and then the recording process is repeated. It typically takes 20 seconds to record 300 holograms, which corresponds to 67 ms for recoding a single hologram. From each measured hologram, the optical field was retrieved using a field retrieval algorithm [33]. The retrieved fields at the same wavelength were then used to reconstruct the 3-D RI tomogram using the optical diffraction tomography algorithm with the first order Rytov approximation [34-37]. The resulting missing information due to the limited acceptance angle of the optical system is filled by using the non-negativity iterative algorithm [38, 39]. Detailed algorithms and used MatLab codes for reconstructing tomography using Rytov approximation and non-negativity interation could be found here [28].

Theoretically, the lateral and axial resolution of optical diffraction tomography is defined as $\lambda/4NA$ and $\frac{1}{2}\frac{\lambda}{n_{medium}-\sqrt{n_{medium}^2-NA^2}}$, respectively, where $n_{medium}$ is the RI of a used medium [35]. These resolutions are linearly proportional to the used wavelength; the theoretical lateral and axial resolution is 181 and 978 nm at $\lambda$ = 580 nm, respectively. The lateral and axial resolution was experimentally measured by measuring a 3–μm–diameter microsphere and analyzing the profile along the edge of the microsphere, which was measured as 296 nm and 1.12 μm ($\lambda$ = 580 nm), respectively. The discrepancy between the theoretical and experimental resolution might be due to a slight defocus effect induced by a residual chromatic aberration. Resolution could be further improved by using an objective lens with higher NA and scanning illumination angle more densly.

To verify the capability of HS-ODT, we first measured the spectroscopic tomograms of polystyrene beads (3 μm diameter, 79166, Sigma-Aldrich Inc., USA) submerged in immersion oil. For five individual polystyrene bead, 10 RI tomograms were measured for various wavelengths ranging from 480 nm to 710 nm. The RI tomograms of a bead at three representative wavelengths are shown in Fig. 2. The reconstructed 3-D RI distributions of the polystyrene beads clearly show the spherical shape of the bead with the correct RI values. The measured RI values are in good agreement with those of polystyrene [40], over a broad wavelength range [Fig. 2(d)]. The error between the measured and expected RI values was 4.1 ± 3.3%. The error thought to be mainly resulted from missing angular information due to the limited NA of the objective lens. Appropriate numerical algorithms for filling missing information could improve RI accuracy [39]. However, it should also be noted that the reported RI values of polystyrene varies depending on measurement techniques and curing conditions [41].

To demonstrate the applicability of the present method for measuring the 3-D RI dispersion of biological samples, we next measured the hyperspectral RI tomograms of human RBCs. RBCs extracted from a healthy donor were diluted in a phosphate buffered saline (PBS) solution and sandwiched between two coverslips. The same imaging and reconstruction procedures as above were performed. 3-D RI tomograms of 5 RBCs were measured at 12 different wavelengths ranging from 480 nm to 680 nm.

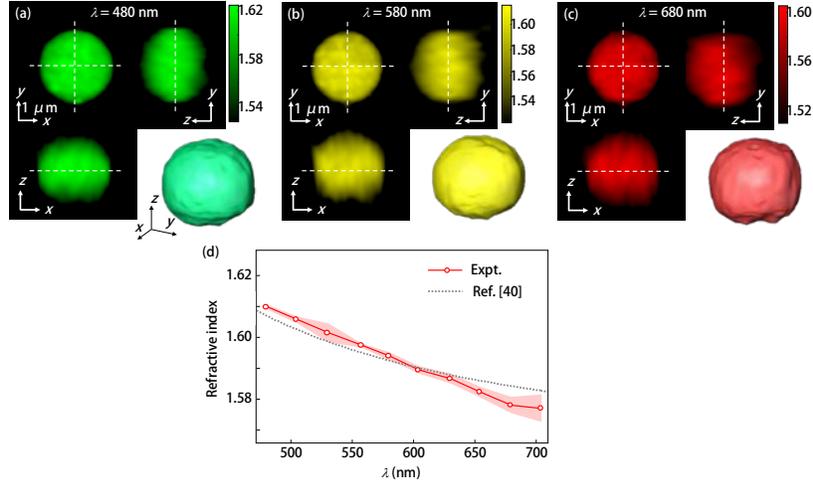

Fig. 2. Cross-sections of the 3-D RI tomograms and the rendered isosurfaces of a polystyrene bead measured at a wavelength of (a) 480, (b) 580, and (c) 680 nm. (d) The averaged RI dispersion of the polystyrene beads. The shaded area represents standard deviations of 5 measurements. Previously reported RI values of polystyrene from Ref. [40] are also plotted in a dotted line for comparison purposes.

The representative 3-D RI tomograms at three different wavelengths are shown in Figs. 3(a-c). The distinct biconcave shapes of the RBCs are clearly seen in the figure, and the measured RI values of the RBCs are in a range consistent with previous reports [42-44]. Because RBCs have a homogeneous distribution of RI, the hemoglobin (Hb) concentration of the RBC cytoplasm was calculated from the measured RI values. The measured average value for the Hb concentration of the RBCs was 34.5 ± 1.9 g/dL, which is within the physiological range of healthy RBCs. The exact RI values of RBCs vary over a range because the physiological level of the Hb concentration is wide; RI values of 1.38–1.39 at 532 nm corresponds to a Hb concentration of 28−36 g/dL.

To further analyze the measured RI dispersion of Hb, the values of the RI difference between the RBCs and, the medium $\Delta n(\lambda)$ was normalized by the mean RI difference $\langle \Delta n(\lambda) \rangle_\lambda$ in the measured wavelength range as follows:

$$\frac{\Delta n(\lambda)}{\langle \Delta n(\lambda) \rangle_\lambda} = \frac{\alpha(\lambda) C_{Hb}}{\langle \alpha(\lambda) C_{Hb} \rangle_\lambda} = \frac{\alpha(\lambda)}{\langle \alpha(\lambda) \rangle_\lambda} \; , \quad (1)$$

where $\alpha$ and $C_{Hb}$ are the RI increment and the concentration of Hb, respectively. As foremementioned, Hb concentrations and thus RI values of individual RBCs varies from cell to cell. However, Hb increment is an intrinsic molecular property of Hb, and does not vary in different RBCs. With this normalization, therefore, the cell-to-cell variation of the RI values was eliminated, and the dispersive characteristics of the $\alpha(\lambda)$ of Hb molecule can clearly be seen [Fig. 3(d)], which are in good agreement with reported values [43].

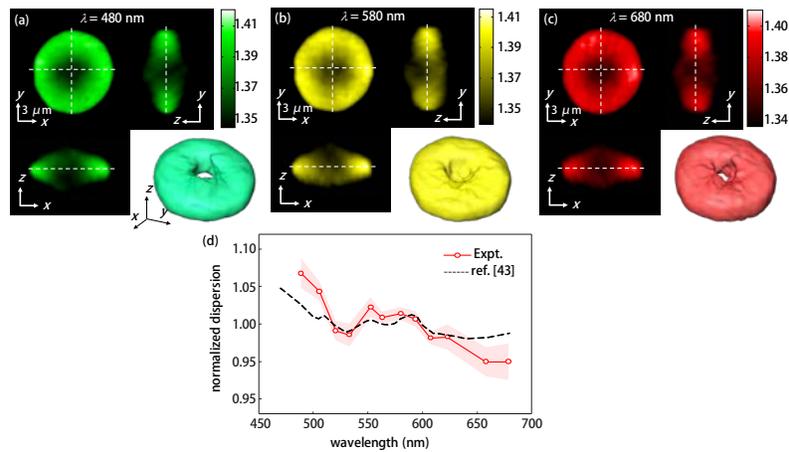

Fig. 3. Cross-sections of the 3-D RI tomograms and the rendered isosurfaces of a RBC from a healthy donor at wavelength of (a) 480, (b) 581, and (c) 658 nm. (d) The dispersive characteristics of the normalized RI values of 5 RBCs. The shaded area represents the standard deviation. Previously reported values [43] are also plotted in a dotted line for comparison purposes

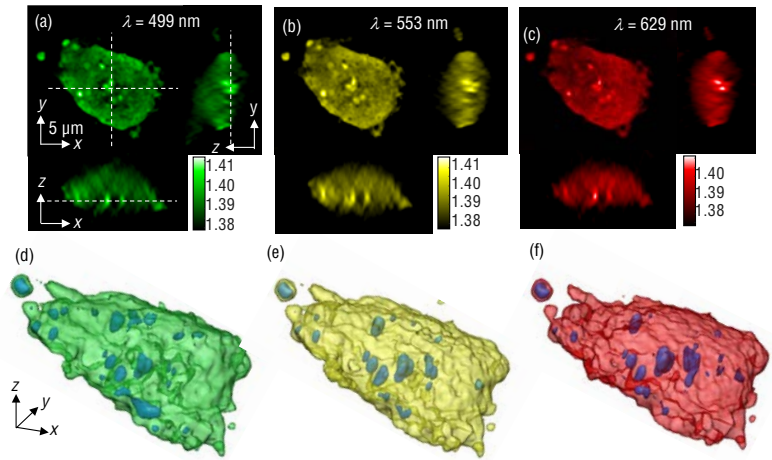

Fig. 4. Cross-sections of the 3-D RI tomograms and the rendered isosurfaces of a Huh7 cell at wavelength of (a) 499 nm, (b) 553 nm, and (c) 629 nm. (d-f) Blue colored isosurfaces inside the cell represent subcellular organelles presumed to be vesicles.

To further demonstrate the applicability of HS-ODT, we measured the hyperspectral 3-D RI maps of eukaryotic cells with complex internal structures. Hepatocyte cells (Huh-7 cell line, Apath, USA) were prepared according to a standard protocol [45]. Briefly, Huh-7 cells were maintained in Dulbecco's Modified Eagle Medium (DMEM, Gibco, USA) supplemented with 10% heat-inactivated fetal bovine serum, 4500 mg/L D- glucose, L-glutamine, 110 mg/L sodium pyruvate, sodium bicarbonate, 100 U/mL penicillin and 100 μg/mL streptomycin. The cells were sub-cultivated for 4 hours before the experiments, and then diluted in PBS solution before the measurements. To prevent the movement of subcellular structures during the measurements, the hepatocyte cells were fixed with paraformaldehyde which has been known to preserve internal structure of cells compared to othe fixation methods [46]. Briefly, the cells were rinsed 3 times with 4°C PBS and then fixed in PBS containing 4% paraformaldehyde for 30 minutes at room temperature. Next, the cells were rinsed 3 times with distilled water and then with mounting solution (DAKO Corp., USA).

The 3-D RI tomograms of Huh-7 were measured at 10 different wavelengths in the visible range. RI tomograms at three representative wavelengths are shown in Fig. 4. The tomograms clearly show the overall shapes as well as the internal structures of the hepatocytes (Fig. 4). In particular, subcellular organelles including nuclei and vesicles are clearly visualized and exhibit different wavelength dispersions.

## 3. Discussion and conclusion

Herein, we present a technique for measuring the hyperspectral 3-D RI tomograms of microscopic objects. Exploiting wavelength-scanning illumination and the common-path 3-D interferometric microscopy technique, 3-D RI maps of a sample at a specific wavelength can be precisely measured with the diffraction limited spatial resolution. The capability of the present method was verified by measuring the RI tomograms of polystyrene beads, human RBCs, and hepatocytes at multiple wavelengths. The measured hyperspectral RI tomograms clearly show the morphology of the samples as well as the RI values over a wide range of wavelengths.

Although the concept is demonstrated in the visible wavelengths in this paper, the present method can be readily applied to other ranges of wavelengths. By using the proper light sources, lenses, and camera, the spectral range can be extended or shifted to other spectral ranges such as the ultra-violet and infra-red (IR). For example, because the used supercontinuum laser generates light with wavelengths of 450 – 2400 nm, the spectral range of the system could be expanded by combining lenses and a camera for the IR region as well.

Currently, the total acquisition time for measuring one RI tomogram at a specific wavelength is approximately 20 sec.; measurements of 300 2-D holograms are required to reconstruct one 3-D tomogram. This speed is mainly limited by the spectral density of the light source. The acquisition time can be reduced by using a high-power source. The total acquisition time also depends on the number of 2-D holograms for reconstructing a single tomogram and on the illuminating wavelengths. Both can be adjusted to suit the purposes of experiments and experimental conditions. For instance, for samples with simple geometry, 10 2-D holographic measurements can provide a high quality 3-D tomogram, which suggests the acquisition time could be reduced by a factor of 30 [47].

Hyperspectral RI tomography has potential for label-free 3-D imaging of biological samples through the unique advantage of measuring the 3-D RI distribution at various wavelengths. We expect that hyperspectral RI dispersion measurements will provide useful molecular information about live cells and tissue via optical dispersion of intrinsic moleculesm, which can be used complimentarily along with existing fluorescence imaging techniqeus. The measured 4-D data set, the spectroscopic 3-D RI maps, can be used to provide information on molecular content, exploiting principle component analysis [48]. The present method will find potential applications in which label-free molecular imaging is required.

**Acknowledgments**
This work was supported by the National Research Foundation (NRF) of Korea (2014K1A3A1A09063027, 2013M3C1A3063046, 2012-M3C1A1-048860, 2014M3C1A3052537), APCTP, and KUSTAR-KAIST project.